# Title: Interface-Induced Superconductivity in Magnetic Topological Insulator-Iron Chalcogenide Heterostructures

**Authors:** Hemian Yi[1†], Yi-Fan Zhao[1†], Ying-Ting Chan[2†], Jiaqi Cai[3†], Ruobing Mei[1], Xianxin Wu[4], Zi-Jie Yan[1], Ling-Jie Zhou[1], Ruoxi Zhang[1], Zihao Wang[1], Stephen Paolini[1], Run Xiao[1], Ke Wang[5], Anthony R. Richardella[1,5], John Singleton[6], Laurel E. Winter[6], Thomas Prokscha[7], Zaher Salman[7], Andreas Suter[7], Purnima P. Balakrishnan[8], Alexander J. Grutter[8], Moses H. W. Chan[1], Nitin Samarth[1,5,9], Xiaodong Xu[3,10], Weida Wu[2]*, Chao-Xing Liu[1]*, and Cui-Zu Chang[1,5]*

**Affiliations:**

[1] Department of Physics, The Pennsylvania State University, University Park, PA 16802, USA

[2] Department of Physics and Astronomy, Rutgers University, Piscataway, NJ 08854, USA

[3] Department of Physics, University of Washington, Seattle, WA 98195, USA

[4] CAS Key Laboratory of Theoretical Physics, Institute of Theoretical Physics, Chinese Academy of Sciences, Beijing 100190, China

[5] Materials Research Institute, The Pennsylvania State University, University Park, PA 16802, USA

[6] National High Magnetic Field Laboratory, Los Alamos, New Mexico 87544, USA

[7] Laboratory for Muon Spectroscopy, Paul Scherrer Institute, 5232 Villigen PSI, Switzerland

[8] NIST Center for Neutron Research, National Institute of Standards and Technology, Gaithersburg, Maryland 20899, USA

[9] Department of Materials Science and Engineering, The Pennsylvania State University, University Park, PA 16802, USA

[10] Department of Materials Science and Engineering, University of Washington, Seattle, WA 98195, USA.

*Correspondence to: wdwu@physics.rutgers.edu (W. W.); cxl56@psu.edu (C.-X. L.); cxc955@psu.edu (C.Z.C.).

† These authors contributed equally to this work.

**Abstract:** When two different electronic materials are brought together, the resultant interface often shows unexpected quantum phenomena, including interfacial superconductivity and Fu-Kane topological superconductivity (TSC). Here, we use molecular beam epitaxy (MBE) to synthesize heterostructures formed by stacking together two magnetic materials, a ferromagnetic topological insulator (TI) and an antiferromagnetic iron chalcogenide (FeTe). We discover emergent interface-induced superconductivity in these heterostructures and demonstrate the trifecta occurrence of superconductivity, ferromagnetism, and topological band structure in the magnetic TI layer, the three essential ingredients of chiral TSC. The unusual coexistence of ferromagnetism and superconductivity can be attributed to the high upper critical magnetic field that exceeds the Pauli paramagnetic limit for conventional superconductors at low temperatures. The magnetic TI/FeTe heterostructures with robust superconductivity and atomically sharp interfaces provide an ideal wafer-scale platform for the exploration of chiral TSC and Majorana physics, constituting an important step toward scalable topological quantum computation.

**One-Sentence Summary:** We demonstrated interface-induced superconductivity formed by stacking together a ferromagnetic topological insulator and an antiferromagnetic iron chalcogenide.

**Main Text:** Superconductivity is a fascinating macroscopic quantum phenomenon in which electrons travel without resistance. The vanishing resistivity is due to the spontaneous condensation of bosonic Cooper pairs (*1, 2*). Besides natural compounds, superconductivity also emerges in artificially constructed heterostructures such as $LaAlO_3/SrTiO_3$ (*3, 4*) and $La_{1.55}Sr_{0.45}CuO_4/La_2CuO_4$ (*5*) wherein the parent materials themselves are not superconductors. Moreover, an unusual form of superconductivity known as topological superconductivity (TSC) has been proposed to emerge in heterostructures formed by a material with strong spin-orbit coupling (SOC) coupled with *s*-wave superconductivity (*6-8*). Over the past decade, the possibility of realizing TSC in such heterostructures has attracted significant attention, driven by the possibility that its excitations [i.e. Majorana zero modes (MZMs)] can be used for fault-tolerant topological quantum computation (*9, 10*).

Topological insulators (TIs) are materials with extremely strong SOC, in which the interior is insulating but electrons can travel freely along the edges/surfaces (*11, 12*). These qualities make TIs a natural platform to pursue TSC (*13*). However, trapping MZMs in a TI/superconductor heterostructure requires the external breaking of time-reversal symmetry. This requirement often inherently suppresses superconductivity and destroys the TSC phase. Magnetic TI/superconductor heterostructures offer an alternative platform for creating the (chiral) TSC phase and MZMs at zero magnetic field. This scheme exploits the quantum anomalous Hall (QAH) effect to realize chiral Majorana modes propagating along the 1D boundaries of the sample (*14-16*). This approach allows for constructing Majorana interferometry that provides both a new probe of coherent Majorana physics and allows for the readout of topological qubits (*17, 18*). The realization of the QAH effect in molecular beam epitaxy (MBE)-grown magnetic TI, specifically Cr- and/or V-doped $(Bi, Sb)_2Te_3$ films/heterostructures (*19*), now provides a well-controlled material basis for designing a TSC platform derived from magnetic TI/superconductor heterostructures.

To date, QAH/superconductor hybrid structures have been fabricated by sputtering a thick *s*-wave superconductor (e.g. ~100 nm Nb) layer on magnetic TI films (*16*). These superconducting films are usually polycrystalline which weakens their proximity effect across the QAH/superconductor interface (*16*). QAH/superconductor heterostructures with atomically sharp interfaces are highly desirable for the exploration of chiral TSC. Moreover, in practice, the superconductivity in MBE-grown thin films is usually suppressed or disappears once a magnetic layer is grown on top, primarily owing to the effective pair-breaking mechanism of spin-flip scattering at the interface (*20*). As an alternative, QAH/non-superconducting material heterostructures with interfacial superconductivity that host chiral TSC being pursued. This arrangement fulfills the three essential ingredients of chiral TSC, i.e., superconducting, ferromagnetic, and topological orders(*14*). Recently, superconductivity has been found in $Bi_2Te_3/FeTe$ and $Sb_2Te_3/FeTe$ heterostructures with the zero resistance superconducting temperatures $T_{c,0}$ ~11.5 K and ~3.1 K, respectively (*21, 22*). $(Bi/Sb)_2Te_3$ is a prototype three-dimensional (3D) TI (*23-25*), while FeTe is a non-superconducting antiferromagnetic iron chalcogenide (*26-28*). Moreover, since $(Bi/Sb)_2Te_3$ with and without Cr doping share the same lattice structure and similar growth recipes, this provides an opportunity to synthesize the Cr-doped $(Bi, Sb)_2Te_3$ films (*19*) on an antiferromagnetic FeTe layer to form Cr-doped $(Bi, Sb)_2Te_3/FeTe$ heterostructures (Fig. 1A).

In this work, we use MBE to fabricate heterostructures formed by $m$ quintuple layers (QL) ferromagnetic TI Cr-doped (Bi, Sb)$_2$Te$_3$ and $n$ unit cells (UC) antiferromagnetic iron chalcogenide FeTe films. Below we denote $m$QL Cr-doped (Bi, Sb)$_2$Te$_3$/$n$UC FeTe as the ($m$,$n$) heterostructure. We find that these Cr-doped (Bi, Sb)$_2$Te$_3$/FeTe heterostructures have an atomically sharp interface. By performing electrical transport, reflective magnetic circular dichroism (RMCD), magnetic force microscopy (MFM), scanning tunneling microscopy and spectroscopy (STM/S), low energy muon spin relaxation (LE-µSR), and angle-resolved photoemission spectroscopy (ARPES) measurements (*29*), we reveal and characterize emergent interface-induced superconductivity in Cr-doped (Bi, Sb)$_2$Te$_3$/FeTe heterostructures and demonstrate the trifecta occurrence of superconductivity, ferromagnetism, and topological order in the Cr-doped (Bi, Sb)$_2$Te$_3$ layer. We also find the upper critical magnetic field of the emergent superconductivity to be very high (>40 T), which may be responsible for the robust superconductivity in an entirely magnetic heterostructure that interfaces a ferromagnet with an antiferromagnet.

We first characterize the Cr-doped (Bi, Sb)$_2$Te$_3$/FeTe heterostructures. Despite the different lattice structures (hexagonal vs cubic), both layers show sharp and streaky "1×1" reflection high-energy electron diffraction (RHEED) patterns (Fig. S1), presumably due to the hybrid symmetry epitaxy (*30*). X-ray diffraction (XRD) spectra show peaks of both layers, further confirming the high crystalline property of the Cr-doped (Bi, Sb)$_2$Te$_3$/FeTe heterostructures (Fig. S2). Cross-sectional scanning transmission electron microscopy (STEM) measurements reveal an atomically sharp interface across Cr-doped (Bi, Sb)$_2$Te$_3$ and FeTe (Figs. 1B and S3). The QL structure of the Cr-doped (Bi, Sb)$_2$Te$_3$ layer and the trilayer structure of the FeTe layer are seen. To examine the topological band structure, we perform *in-vacuo* ARPES measurements on an (8,50) heterostructure (Fig. 1C). Linearly-dispersed Dirac surface states crossing the chemical potential are observed, consistent with the band spectra of the QAH samples at room temperature (*19, 31*). Indeed, we observe the well-quantized QAH effect in an 8 QL Cr-doped (Bi,Sb)$_2$Te$_3$ layer with the same components on SrTiO$_3$(111) substrates (Fig. S4).

Next, we perform *ex-situ* transport measurements on a series of Cr-doped (Bi,Sb)$_2$Te$_3$/FeTe heterostructures. Figure 1D shows the temperature dependence of the sheet longitudinal resistance (i.e. $R$-$T$ curves) of ($m$,20) heterostructures. Although the 20UC FeTe film grown on SrTiO$_3$(100) shows a non-superconducting antiferromagnetic behavior (Fig. S5), superconductivity appears after the deposition of 1 QL Cr-doped (Bi,Sb)$_2$Te$_3$ layer, showing clear evidence of interface-triggered superconductivity. For the $m$=1 sample, the onset $T_{c,\text{onset}}$ and the zero resistance $T_{c,0}$ are ~9.6 K and ~2.9 K, respectively. With increasing $m$. $T_{c,\text{onset}}$ increases slightly but $T_{c,0}$ increases much more before showing saturation for $m \geq 4$. For the $m$=8 sample, $T_{c,\text{onset}}$ and $T_{c,0}$ appear to show optimal values of ~12.4 K and ~10.0 K, respectively (Fig.1D).

We then stay with $m$=8 and systematically study the evolution of the emergent superconductivity as a function of the number of FeTe layer $n$ (Fig. 1E). We find that the $n \geq 20$ samples show sharp superconducting transitions with $T_{c,0}$ ranging between 10.0 K and 11.4 K. When $n$ is reduced from 20, the superconductivity is suppressed and the superconducting transition window broadens, indicating enhanced phase fluctuations possibly due to spatial confinement of the Cooper pairs (*32*). For the $n$ =18 sample, $T_{c,\text{onset}}$ ~12.2 K and $T_{c,0}$ ~2.0 K (Fig.1E). For the $n \leq$ 16 samples, the zero-resistance state fades away but the hump feature near $T_{c,\text{onset}}$ persists, indicating the formation of local superconducting order without long-range phase coherence. For $T < T_{c,\text{onset}}$, the $n$ = 16 sample shows metallic behavior, while for the $n \leq 14$ samples, $R$ first drops and then increases to a few kΩ (Fig.1E). The upturn feature in $n \leq 14$ samples indicates the presence of an insulating phase in the low-temperature regime. We note that the observed superconductor-metal-insulator phase transition for $n \leq 16$ is analogous to that found in patterned cuprate superconductors, in which the nearly-linear $R$-$T$ curve before entering the zero-resistance state has been attributed to the disorder-induced Bosonic anomalous metal phase (*33*).

To confirm the existence of ferromagnetism, we first perform electrical transport measurements on the Cr-doped (Bi, Sb)$_2$Te$_3$/FeTe heterostructures. Figure 2A shows the $R$-$T$ curve of the (8,50) heterostructure from room temperature down to $T$=2 K. Its $T_{c,onset}$ is ~12.4 K and $T_{c,0}$ is ~11.4 K. Interestingly, we observe a broad semiconducting-to-metallic crossover near $T$ ~60 K. Such behavior in the $R$-$T$ curve of FeTe has been associated with the paramagnetic-to-antiferromagnetic phase transition (Fig. S5) (*22*). Figure 2B shows the Hall traces near the superconducting transition regime. At $T$ =10 K, which is below $T_{c,0}$ ~ 11.4 K, the Hall resistance $R_{yx}$ value is zero. The vanishing value of $R_{yx}$ here is a result of the zero-resistance state of the superconducting phase. At $T$=12 K, which is located in the superconducting transition regime (i.e. $T_{c,0} < T < T_{c,onset}$), a small hysteresis loop appears, implying the existence of ferromagnetism. By further increasing $T$, the sample shows a hysteresis loop with larger $R_{yx}$ at $T$ =14 K, further confirming the existence of the ferromagnetism in Cr-doped (Bi, Sb)$_2$Te$_3$/FeTe heterostructure above its superconducting $T_{c,onset}$ (Fig. 2B).

To demonstrate the coexistence of superconductivity and ferromagnetism at $T < T_{c,0}$, we perform RMCD measurements on the same (8,50) heterostructure. Figure 2C shows the magnetic field $\mu_0H$ dependence of the RMCD signal at $2\ K \leq T \leq 14\ K$. Hysteresis loops are observed in the entire temperature range, demonstrating the persistence of ferromagnetism in the superconducting regime. For $T > T_{c,0}$, the observed hysteresis loops and coercive field $\mu_0H_c$ agree well with the value deduced from the Hall measurements at the same temperature (Figs. 2B and 2C). Similar behavior is observed in the (8,20) FeTe heterostructure (Fig. S6).

The coexistence of superconductivity and ferromagnetism is further supported by direct visualization of ferromagnetic domains using MFM with *in-situ* transport measurements. Figures 3A to 3H show the MFM images of magnetic domains in the (8,20) heterostructure at different $\mu_0H$ and $T$=2.2 K. The sample is field cooled ($\mu_0H$ ~ −0.01 T) through the Curie temperature ($T_{Cuire}$ ~20 K) to ensure a single magnetic domain with downward magnetization, as shown by the uniform MFM domain (Fig. 3A). The superconducting state is confirmed by the zero resistance from *in-situ* transport measurements (Fig. S6). By sweeping $\mu_0H$ upward from -0.01 T, speckles of reversed magnetic domains (blue regions in Fig. 3B) randomly nucleate at $\mu_0H$ = 0.047 T. As $\mu_0H$ increases further, the blue regions (i.e. the upward magnetic domains) expand at the expense of the red regions (i.e. the downward magnetic domains), gradually filling up the entire scan area (Figs. 3C to 3F). At $\mu_0H$ = 0.09 T, the MFM image is uniformly blue, corresponding to a uniform upward magnetization (Fig. 3H). The evolution of domain nucleation and domain wall propagation in the zero-resistance state demonstrates the coexistence of long-range ferromagnetic order and the emergent superconducting state (*34*).

The direct observation of magnetic domains also allows us to construct the ferromagnetic hysteresis loop. Figure 3I shows the $\mu_0H$ dependence of the normalized magnetization $M/M_s$ ($M_s$ is the saturated magnetization) at $T$ =2.2 K, which is estimated from each MFM image using a histogram analysis(*34*). The upward and downward domains are equally populated when $M/M_s$ ~0 at coercive field $\mu_0H_c$ ~0.061 T. This is in good agreement with the peak position of the root-mean-square (RMS) value $\delta f$ of the MFM signal (Fig. 3I) and the value obtained from the RMCD measurement at $T$=2 K (Fig. 3J). In addition, similar magnetic domain behaviors are observed in the normal state at $T > T_{c,onset}$ ~ 12.4 K (Fig. S7), demonstrating that the ferromagnetic order is not affected by the emergence of the superconducting state.

To confirm and pinpoint the interface-induced superconductivity and the proximity effect in the ferromagnetic Cr-doped (Bi,Sb)$_2$Te$_3$ layer, we perform low-temperature STM/S measurements on ($m$, 38) heterostructures of different $m$ (Figs. 4 and S8 to S13). For the 38UC FeTe layer that is partially covered by 1 QL Cr-doped (Bi, Sb)$_2$Te$_3$ (i.e., the $m$<1 sample, Fig. S8A), the $dI/dV$ spectra across the interface between 1 QL Cr-doped (Bi, Sb)$_2$Te$_3$ and 38UC FeTe layers show the presence of the superconducting gap on the surface of 1 QL Cr-doped (Bi, Sb)$_2$Te$_3$ but absent on the FeTe surface at $T$=310 mK (Fig. S8B). This is consistent with our transport data (Figs. 1D and S5). We note that the superconducting gap is spatially uniform and homogeneous on the surface of the $m$=1 sample. This observation directly confirms the emergence of the interface-induced superconductivity in Cr-doped (Bi, Sb)$_2$Te$_3$/FeTe heterostructures.

Next, we focus on probing the proximity-induced superconducting gap on the top surface of the ferromagnetic Cr-doped (Bi,Sb)$_2$Te$_3$ layer with different thicknesses. Figure 4A shows the large-scale STM image of the (8, 38) heterostructure. The surface of the Cr-doped (Bi,Sb)$_2$Te$_3$ layer exhibits spiral structures, which are characterized by atomically flat terraces with a width of a few tens of nanometers (*19, 35*). The height of the terrace is ~1 nm, corresponding to the thickness of 1QL TI. The STM topographic image with atomic resolution (Fig. 4B) shows randomly distributed triangle-shaped dark features on the surface, which are Cr substitutions at Bi/Sb sites in the Cr-doped (Bi,Sb)$_2$Te$_3$ layer (*19*). Figure 4C shows the $dI/dV$ spectra measured on the ($m$, 38) heterostructures with different $m$ at $T$=310 mK. The suppressed spectral weight around the Fermi level signifies a superconducting gap on the top surface of the Cr-doped (Bi, Sb)$_2$Te$_3$ layer with $m$ up to 10, suggesting a strong proximity effect. The gap feature can be fitted with the Dynes formula (Fig. S9)(*36*). The resultant gap size is in good agreement with that of undoped TI/FeTe heterostructures in prior studies (*37*). With increasing $m$, the size of the superconducting gap gradually decreases (Figs. 4C and 4D). This behavior resembles that found in undoped TI/FeTe heterostructures (*37*). Moreover, the superconducting gap shrinks with increasing temperature and disappears at $T$ ~6.5 K (Figs. 4E and S10), slightly lower than the superconducting transition temperature found in transport measurements. This further confirms that the gap observed here is the proximity-induced superconducting gap on the top surface of the Cr-doped (Bi, Sb)$_2$Te$_3$ layer. Combining transport, RMCD, MFM, and STM/S results (Figs. 2 to 4), we thus unambiguously demonstrate the coexistence of superconductivity and ferromagnetism in the Cr-doped (Bi, Sb)$_2$Te$_3$ layer, indicating an unconventional mechanism of superconductivity. Furthermore, our LE-µSR results suggest that the superconductivity does not suppress the magnetic order but instead occurs commensurately with a phase transition to a more uniform, ordered magnetic state centered at the Cr-doped (Bi, Sb)$_2$Te$_3$/FeTe interface (Figs. S17 to S19).

To investigate the physical mechanism responsible for the coexistence of superconductivity and ferromagnetism and the superconducting pair-breaking mechanism, we vary the angle θ between the normal direction of the film and the direction of µ$_0$H and measure the values of the upper critical magnetic field µ$_0$H$_{c2}$ at $T$=1.5 K (Fig. S14). We find µ$_0$H$_{c2}$ to be independent of θ, thus confirming the superconductivity to be isotropic. We note that the large and isotropic µ$_0$H$_{c2}$~44.5 T at $T$=1.5 K exceeds the Pauli paramagnetic limit of µ$_0$H$_p$ =1.84$T_c$ ~23.0 T for a conventional $s$-wave superconductor. This large µ$_0$H$_{c2}$ may be responsible for the coexistence of ferromagnetism and superconductivity in Cr-doped (Bi, Sb)$_2$Te$_3$/FeTe heterostructures since the pair-breaking mechanism from exchange coupling of ferromagnetism is similar to the Pauli paramagnetic effect (*38*). Moreover, the µ$_0$H$_{c2}$ ~$T/T_c$ phase diagrams near $T_c$ are fitted well with the Ginzburg–Landau (GL) theory in the 2D limit (Figs. S15 and S16). The extracted in-plane GL coherence length is ~2.2 nm and the effective thickness of the superconducting region is ~10.4 nm for the (8,38) heterostructure (Figs. S15). The latter value is less than the thickness of the FeTe layer (~24.7 nm), further implying that the superconductivity realized in Cr-doped (Bi, Sb)$_2$Te$_3$/FeTe heterostructures stems from an interfacial effect. Similar values of coherence length and effective superconducting thickness are also found in the (4,20) heterostructure (Figs. S16) with the minimum thickness of the FeTe layer to achieve the zero-resistance state of superconductivity (Fig. 1E), corroborating the interfacial effect.

The observed anisotropic $\mu_0 H_{c2}$ near $T_c$ but isotropic at low temperatures resembles that of iron chalcogenide crystals and thin films (*39-41*), suggesting that the superconductivity in our heterostructures originates from the FeTe layer when it is proximally coupled to the ferromagnetic Cr-doped (Bi, Sb)$_2$Te$_3$ layer. Therefore, the finite resistance in the $n \leq 16$ samples for $T < T_{c,onset}$ is likely a result of spatial confinement-induced phase fluctuations in the FeTe layer, which is induced by its low and inhomogeneous superfluid density (*42, 43*). Although the parent phase of FeTe is an antiferromagnetic metal, superconductivity can emerge after its antiferromagnetic order is weakened by either elemental doping (*26, 27*) or tensile stress(*28*). The hole-type charge carrier transfer effect across the Cr-doped (Bi, Sb)$_2$Te$_3$/FeTe interface might suppress the AFM order, triggering the formation of this interface-induced superconductivity (*44*). This possibility has been suggested by the observation of superconductivity in FeTe films with oxygen adsorption (*45, 46*). The strain effect and RKKY interactions from the spin-momentum locked topological surface states (*47*) coupled with the interfacial FeTe layer may destabilize the antiferromagnetic order in the FeTe layer and thus promote superconductivity (*28*). A more complete understanding of the pairing mechanism requires further studies.

The most striking feature in our experiment is the long superconductivity proximity length in ferromagnetic Cr-doped (Bi, Sb)$_2$Te$_3$ layers up to $m$~10 (Fig. 4D). The charge transfer occurs across the Cr-doped (Bi, Sb)$_2$Te$_3$/FeTe interface induces an asymmetric potential that leads to band bending, thus breaking inversion symmetry in the Cr-doped (Bi, Sb)$_2$Te$_3$ layer (Fig. S20A). Our theoretical calculations show that this asymmetric potential plays a dual role in stabilizing the superconducting proximity effect in the Cr-doped (Bi, Sb)$_2$Te$_3$ layer. First, the asymmetric potential pushes the top surface state so that its energy is closer to the bulk conduction band bottom. This promotes a strong hybridization between the top surface state and the bulk conduction bands, resulting in a very long penetration depth of the top surface state across the entire Cr-doped (Bi, Sb)$_2$Te$_3$ layer (Fig. S20). This enables a strong coupling between the top surface state and superconductivity at the Cr-doped (Bi, Sb)$_2$Te$_3$/FeTe interface. Second, the asymmetric potential breaks the inversion symmetry, leading to the emergence of a mixed pairing state with both singlet and triplet components (*20*). This inversion symmetry breaking removes the energy degeneracy between the top and bottom surface states. Due to the spin-momentum locking of topological surface states, electron spin at the Fermi surfaces becomes predominantly pinned into the 2D plane, while the $z$-directional magnetization only induces a minor tilt in the electron spin (Figs. S21A and S21B). Consequently, the thickness dependence of the proximity-induced superconducting gap exhibits a monotonic decay with a long proximity length (Fig. S21C), in good agreement with our experimental observations (Fig. 4D). Therefore, we demonstrate the existence of long-range superconducting proximity in magnetic TI, even with a spin singlet superconductivity at the magnetic TI/FeTe interface. Given the coexistence of ferromagnetism and superconductivity, our theoretical analysis of the topological phase diagram (Fig. S22) also establishes the MBE-grown Cr-doped (Bi, Sb)$_2$Te$_3$/FeTe heterostructure with an atomically sharp interface as a natural platform to search for the chiral TSC phase (*14*).

To summarize, we use MBE to synthesize Cr-doped (Bi, Sb)$_2$Te$_3$/FeTe heterostructures and find emergent superconductivity in these heterostructures formed by interfacing two magnetic materials. We establish the trifecta occurrence of superconductivity, ferromagnetism, and topological order in these heterostructures and attribute the coexistence of ferromagnetism and robust superconductivity to the high isotropic upper critical magnetic field of the emergent superconductivity. The Cr-doped (Bi, Sb)$_2$Te$_3$/FeTe heterostructures synthesized in this work fulfill all three essential ingredients of chiral TSC (*14*) and thus provide a promising suitable system for the exploration of chiral Majorana physics and the development of a scalable platform for topological quantum computations.

**References and Notes**

1. J. Bardeen, L. N. Cooper, J. R. Schrieffer, Microscopic Theory of Superconductivity. *Phys. Rev.* **106**, 162-164 (1957).
2. J. Bardeen, L. N. Cooper, J. R. Schrieffer, Theory of Superconductivity. *Phys. Rev.* **108**, 1175-1204 (1957).
3. A. Ohtomo, H. Y. Hwang, A high-mobility electron gas at the LaAlO$_3$/SrTiO$_3$ heterointerface. *Nature* **427**, 423-426 (2004).
4. N. Reyren *et al.*, Superconducting interfaces between insulating oxides. *Science* **317**, 1196-1199 (2007).
5. A. Gozar *et al.*, High-temperature interface superconductivity between metallic and insulating copper oxides. *Nature* **455**, 782-785 (2008).
6. J. Alicea, Majorana Fermions in a Tunable Semiconductor Device. *Phys. Rev. B* **81**, 125318 (2010).
7. R. M. Lutchyn, J. D. Sau, S. Das Sarma, Majorana Fermions and a Topological Phase Transition in Semiconductor-Superconductor Heterostructures. *Phys. Rev. Lett.* **105**, 077001 (2010).
8. Y. Oreg, G. Refael, F. von Oppen, Helical Liquids and Majorana Bound States in Quantum Wires. *Phys. Rev. Lett.* **105**, 177002 (2010).
9. V. Mourik *et al.*, Signatures of Majorana Fermions in Hybrid Superconductor-Semiconductor Nanowire Devices. *Science* **336**, 1003-1007 (2012).
10. A. Y. Kitaev, Fault-Tolerant Quantum Computation by Anyons. *Ann. Phys.* **303**, 2-30 (2003).
11. X. L. Qi, S. C. Zhang, Topological Insulators and Superconductors. *Rev. Mod. Phys.* **83**, 1057-1110 (2011).
12. M. Z. Hasan, C. L. Kane, Colloquium: Topological Insulators. *Rev. Mod. Phys.* **82**, 3045-3067 (2010).
13. L. Fu, C. L. Kane, Superconducting Proximity Effect and Majorana Fermions at the Surface of a Topological Insulator. *Phys. Rev. Lett.* **100**, 096407 (2008).
14. X. L. Qi, T. L. Hughes, S. C. Zhang, Chiral Topological Superconductor from the Quantum Hall State. *Phys. Rev. B* **82**, 184516 (2010).
15. J. Wang, Q. Zhou, B. Lian, S. C. Zhang, Chiral Topological Superconductor and Half-Integer Conductance Plateau from Quantum Anomalous Hall Plateau Transition. *Phys. Rev. B* **92**, 064520 (2015).
16. M. Kayyalha *et al.*, Absence of evidence for chiral Majorana modes in quantum anomalous Hall-superconductor devices. *Science* **367**, 64-67 (2020).
17. L. Fu, C. L. Kane, Probing Neutral Majorana Fermion Edge Modes with Charge Transport. *Phys. Rev. Lett.* **102**, 216403 (2009).
18. A. R. Akhmerov, J. Nilsson, C. W. J. Beenakker, Electrically Detected Interferometry of

Majorana Fermions in a Topological Insulator. *Phys. Rev. Lett.* **102**, 216404 (2009).
19. C.-Z. Chang, C.-X. Liu, A. H. MacDonald, Colloquium: Quantum anomalous Hall effect. *Rev. Mod. Phys.* **95**, 011002 (2023).
20. A. I. Buzdin, Proximity effects in superconductor-ferromagnet heterostructures. *Rev. Mod. Phys.* **77**, 935-976 (2005).
21. Q. L. He *et al.*, Two-dimensional superconductivity at the interface of a $Bi_2Te_3$/FeTe heterostructure. *Nat. Commun.* **5**, 4247 (2014).
22. J. Liang *et al.*, Studies on the origin of the interfacial superconductivity of $Sb_2Te_3/Fe_{1+y}Te$ heterostructures. *Proc. Natl. Acad. Sci.* **117**, 221-227 (2020).
23. H. J. Zhang *et al.*, Topological Insulators in $Bi_2Se_3$, $Bi_2Te_3$ and $Sb_2Te_3$ with a Single Dirac Cone on the Surface. *Nat. Phys.* **5**, 438-442 (2009).
24. Y. L. Chen *et al.*, Experimental Realization of a Three-Dimensional Topological Insulator, $Bi_2Te_3$. *Science* **325**, 178-181 (2009).
25. D. Hsieh *et al.*, Observation of Time-Reversal-Protected Single-Dirac-Cone Topological-Insulator States in $Bi_2Te_3$ and $Sb_2Te_3$. *Phys. Rev. Lett.* **103**, 146401 (2009).
26. Y. Mizuguchi, F. Tomioka, S. Tsuda, T. Yamaguchi, Y. Takano, Superconductivity in S-substituted FeTe. *Appl. Phys. Lett.* **94**, 012503 (2009).
27. F. S. Li *et al.*, Interface-enhanced high-temperature superconductivity in single-unit-cell $FeTe_{1-x}Se_x$ films on $SrTiO_3$. *Phys. Rev. B* **91**, 220503 (2015).
28. Y. Han *et al.*, Superconductivity in Iron Telluride Thin Films under Tensile Stress. *Phys. Rev. Lett.* **104**, 017003 (2010).
29. See supplementary materials.
30. X. Yao *et al.*, Hybrid Symmetry Epitaxy of the Superconducting Fe(Te,Se) Film on a Topological Insulator. *Nano Lett.* **21**, 6518-6524 (2021).
31. C. Z. Chang *et al.*, Experimental Observation of the Quantum Anomalous Hall Effect in a Magnetic Topological Insulator. *Science* **340**, 167-170 (2013).
32. V. J. Emery, S. A. Kivelson, Importance of Phase Fluctuations in Superconductors with Small Superfluid Density. *Nature* **374**, 434-437 (1995).
33. C. Yang *et al.*, Signatures of a strange metal in a bosonic system. *Nature* **601**, 205-210 (2022).
34. W. B. Wang *et al.*, Direct Evidence of Ferromagnetism in a Quantum Anomalous Hall System. *Nat. Phys.* **14**, 791-795 (2018).
35. Y. Liu, M. Weinert, L. Li, Spiral Growth without Dislocations: Molecular Beam Epitaxy of the Topological Insulator Bi2Se3 on Epitaxial Graphene/SiC(0001). *Phys. Rev. Lett.* **108**, 115501 (2012).
36. R. C. Dynes, V. Narayanamurti, J. P. Garno, Direct Measurement of Quasiparticle-Lifetime Broadening in a Strong-Coupled Superconductor. *Phys. Rev. Lett.* **41**, 1509-1512 (1978).
37. H. L. Qin *et al.*, Superconductivity in Single-Quintuple-Layer $Bi_2Te_3$ Grown on Epitaxial FeTe. *Nano Lett.* **20**, 3160-3168 (2020).
38. F. S. Bergeret, A. F. Volkov, K. B. Efetov, Odd triplet superconductivity and related phenomena in superconductor-ferromagnet structures. *Rev. Mod. Phys.* **77**, 1321-1373 (2005).
39. M. G. Fang *et al.*, Weak anisotropy of the superconducting upper critical field in $Fe_{1.11}Te_{0.6}Se_{0.4}$ single crystals. *Phys. Rev. B* **81**, 020509 (2010).
40. S. Khim *et al.*, Evidence for dominant Pauli paramagnetic effect in the upper critical field of single-crystalline $FeTe_{0.6}Se_{0.4}$. *Phys. Rev. B* **81**, 184511 (2010).
41. P. Mele, Superconducting properties of iron chalcogenide thin films. *Sci. Technol. Adv. Mater.* **13**, 054301 (2012).

42. D. Chatzopoulos *et al.*, Spatially dispersing Yu-Shiba-Rusinov states in the unconventional superconductor FeTe$_{0.55}$Se$_{0.45}$. *Nat. Commun.* **12**, 298 (2021).
43. D. Cho, K. M. Bastiaans, D. Chatzopoulos, G. D. Gu, M. P. Allan, A strongly inhomogeneous superfluid in an iron-based superconductor. *Nature* **571**, 541-545 (2019).
44. K. Owada *et al.*, Electronic structure of a Bi$_2$Te$_3$/FeTe heterostructure: Implications for unconventional superconductivity. *Phys. Rev. B* **100**, 064518 (2019).
45. W. D. Si *et al.*, Superconductivity in epitaxial thin films of Fe$_{1.08}$Te:O$_x$. *Phys. Rev. B* **81**, 092506 (2010).
46. W. Ren *et al.*, Oxygen Adsorption Induced Superconductivity in Ultrathin FeTe Film on SrTiO$_3$(001). *Materials* **14**, 4584 (2021).
47. Q. Liu, C. X. Liu, C. K. Xu, X. L. Qi, S. C. Zhang, Magnetic Impurities on the Surface of a Topological Insulator. *Phys. Rev. Lett.* **102**, 156603 (2009).
48. C. Z. Chang, Data for "*Interface-Induced Superconductivity in Magnetic Topological Insulator-Iron Chalcogenide Heterostructures*"; https://doi.org/10.7910/DVN/MH1UYX, Harvard Dataverse (2023).

**Acknowledgments:** We are grateful to Yongtao Cui, Liang Fu, Lunhui Hu, Fangsen Li, Andreas Suter, Bing Xia, Di Xiao, Haijun Zhang, and Hao Zheng for helpful discussions and Haiying Wang for STEM sample preparations. **Funding:** This work is primarily supported by the DOE grant (DE-SC0023113), including the MBE growth, ARPES, and electrical transport measurements. The sample characterization is partially supported by the NSF-CAREER award (DMR-1847811) and the Penn State MRSEC for Nanoscale Science (DMR-2011839). The MBE growth and the ARPES measurements are partially performed in the NSF-supported 2DCC MIP facility (DMR-2039351). The STM/S measurements are partially supported by the ARO grant (W911NF2210159). The theoretical calculations are partially supported by the NSF grant (DMR-2241327). The MFM measurements were supported by the DOE grant (DE-SC0018153). The RMCD measurements were supported by the AFOSR grant (FA9550-21-1-0177). Work done at NHMFL is supported by NSF (DMR-1644779 and DMR-2128556) and the State of Florida. C. -Z. C. acknowledges the support from the Gordon and Betty Moore Foundation's EPiQS Initiative (Grant GBMF9063 to C. -Z. C.). Certain commercial equipment, instruments, software, or materials are identified in this paper in order to specify the experimental procedure adequately. Such identifications are not intended to imply recommendation or endorsement by NIST, nor it is intended to imply that the materials or equipment identified are necessarily the best available for the purpose.

**Author contributions:** C. Z. C. conceived and designed the experiment. H. Y. performed the MBE growth, ARPES, and PPMS transport measurements. Y.-F. Z., Z. W., and S. P. performed the STM/S measurements. Y. -T. C. and W. W. performed the MFM measurement. J. C. and X. X. performed the RMCD measurements. J. S., L. E. W., and H. Y. performed the high magnetic field transport measurements. K. W. carried out the TEM measurements. X. W., R. M., and C.-X. L. provided theoretical support. H. Y., Y. -T. C., X. W., W. W., C.-X. L., and C. -Z. C. analyzed the data and wrote the manuscript with input from all authors.

**Competing interests:** The authors declare no competing interests.
**Data and materials availability:** All data in the main text and the supplementary materials are available at (*48*).

**Supplementary Materials:**

Materials and Methods

Supplementary Text

Figures S1-S22

References (*49-70*)

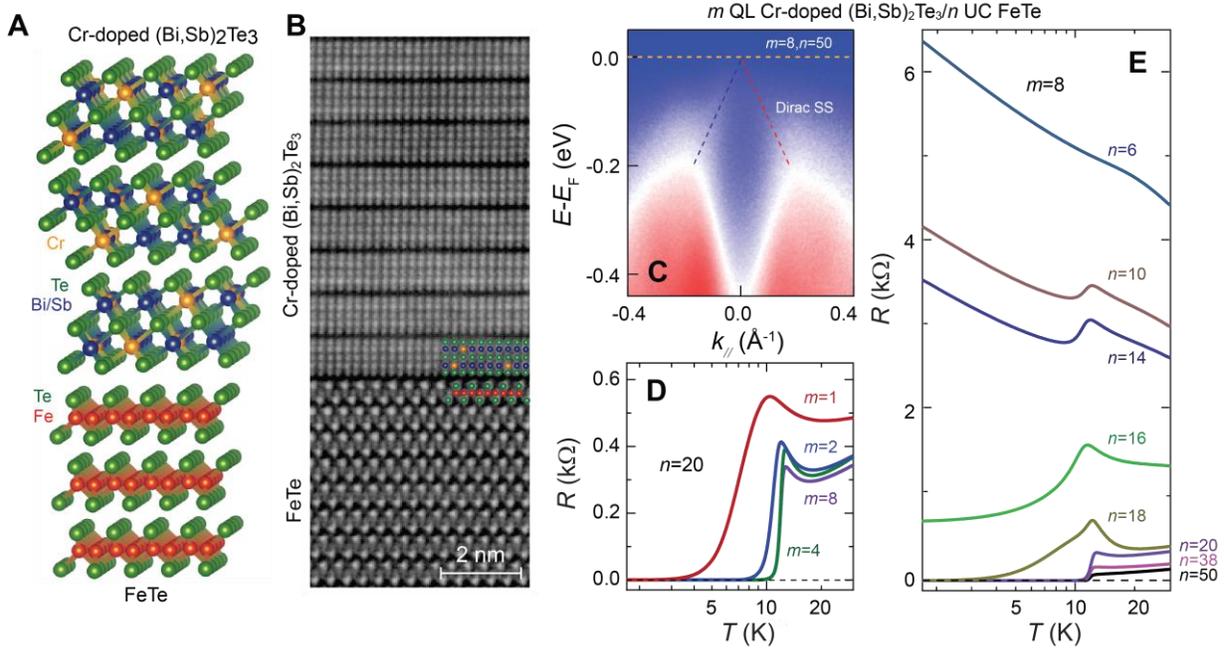

**Fig. 1| Interface-induced superconductivity in Cr-doped (Bi, Sb)$_2$Te$_3$/FeTe heterostructures.**
(**A**) Schematic lattice structure of the Cr-doped (Bi, Sb)$_2$Te$_3$/FeTe heterostructure. (**B**) Cross-sectional STEM image of the (8,50) heterostructures grown on heat-treated SrTiO$_3$ (100) substrate. (**C**) *In-situ* ARPES band map of the (8,50) heterostructure. The Dirac point is located near the chemical potential. The ARPES measurements are performed at room temperature. (**D**) Temperature dependence of the sheet longitudinal resistance $R$ of the ($m$, 20) heterostructures with $m$=1, 2, 4, 8. (**E**) Temperature dependence of $R$ of the (8, $n$) FeTe heterostructures with 6≤$n$≤50.

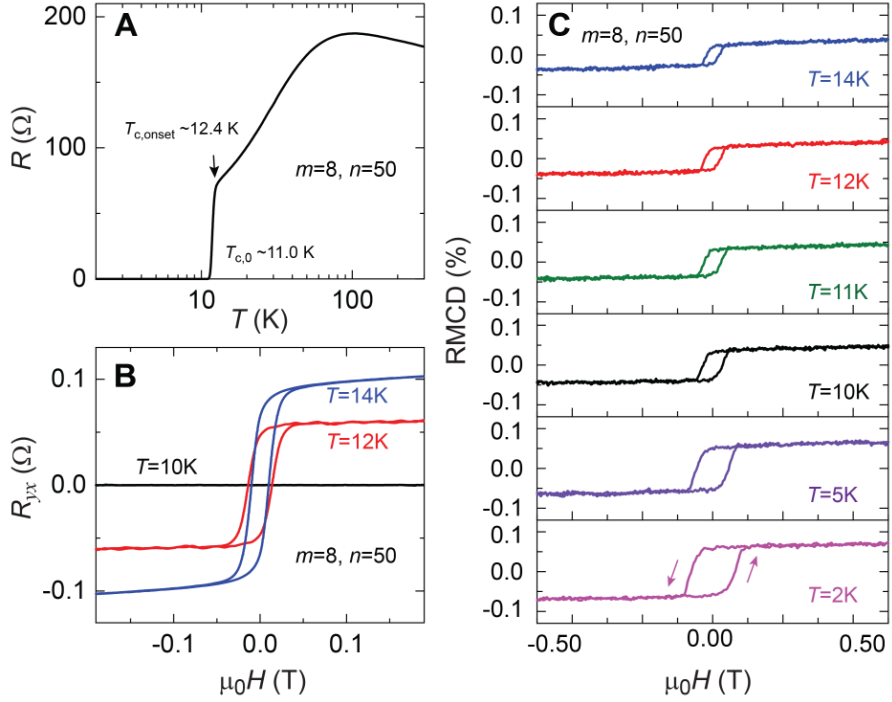

**Fig. 2| Coexistence of ferromagnetism and superconductivity in Cr-doped (Bi, Sb)$_2$Te$_3$/FeTe heterostructures.** (**A**) Temperature dependence of the sheet longitudinal resistance $R$ of the (8,50) heterostructure. (**B**) Magnetic field $\mu_0 H$ dependence of the Hall resistance $R_{yx}$ at $T$=10K (black), $T$=12K (red), and $T$=14K (blue). (**C**) $\mu_0 H$ dependence of the RMCD signal of the (8,50) heterostructure. For $T < T_{c,0}$, the sample shows a hysteresis loop, indicating the coexistence of ferromagnetism and superconductivity.

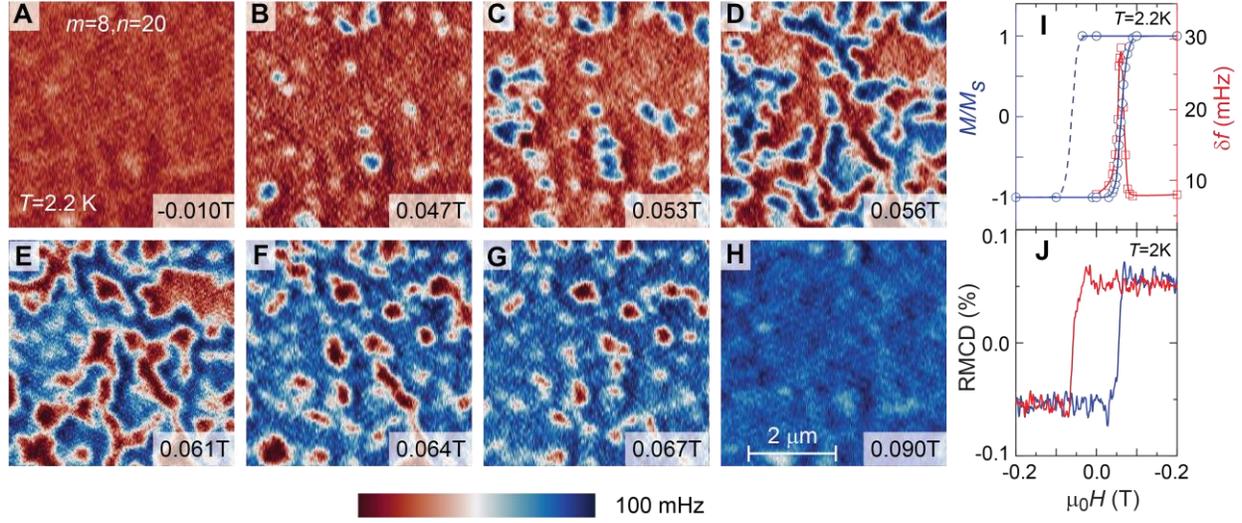

**Fig. 3| Magnetization reversal process in the superconducting state of the Cr-doped (Bi, Sb)$_2$Te$_3$/FeTe heterostructures.** (A-H), MFM images of magnetic domains in the (8,20) heterostructure measured at $\mu_0H$ = -0.01 (A), 0.047T (B), 0.053T (C), 0.056T (D), 0.061T (E), 0.64T (F), 0.067T (G), and 0.090T (H), respectively. All measurements are carried out at $T$ =2.2 K. (I) $\mu_0H$ dependence of the normalized magnetization $M/M_s$ (blue) and the magnetic domain contrast $\delta f$ (red). $M/M_s$ is inferred from each MFM image using histogram analysis. $M_s$ is the saturated magnetization. $\delta f$ is the root-mean-square (RMS) value of each MFM image. (J) $\mu_0H$ dependence of the RMCD signal of the same heterostructure measured at $T$ =2 K.

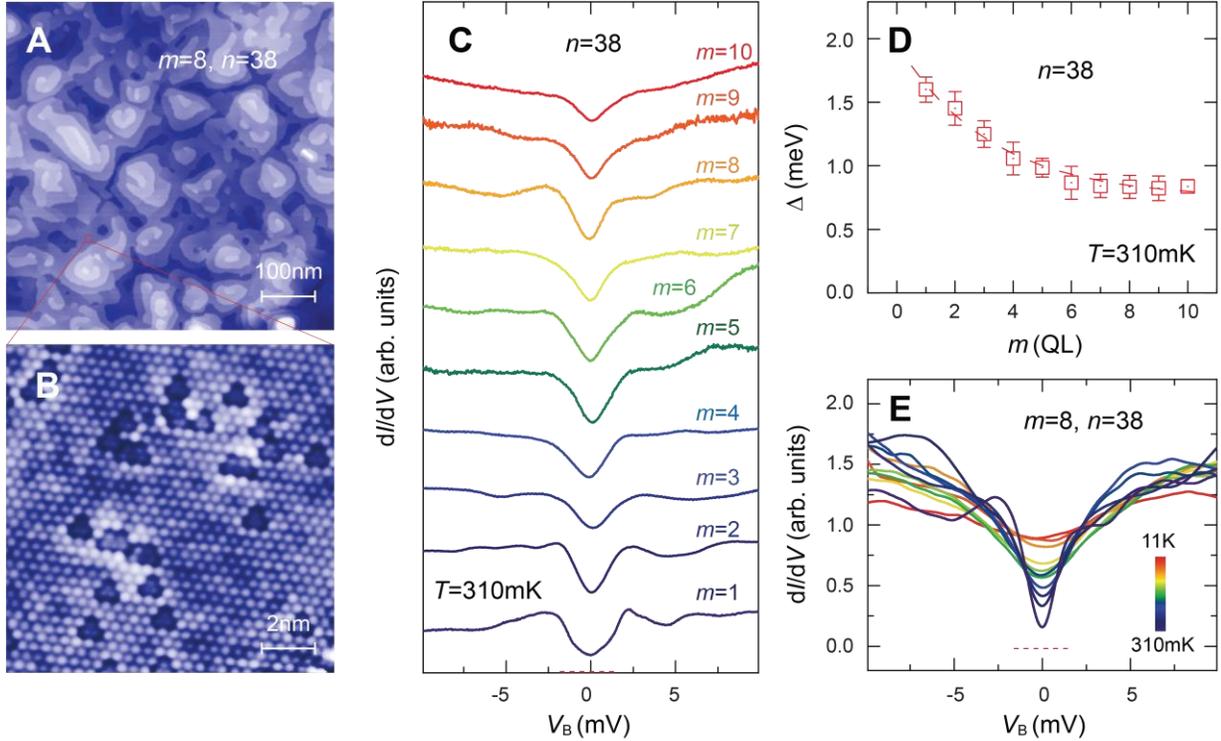

**Fig. 4| STM/S characterization of $m$ QL Cr-doped (Bi, Sb)$_2$Te$_3$/38 UC FeTe heterostructures.**
(**A**) Large-scale STM image of the (8,38) heterostructure showing spiral structures (sample bias $V_B$=+1.5V, tunneling current $I_t$=50pA, $T$=4.2 K). (**B**) Atomic-resolution STM image of the (8,38) heterostructure ($V_B$=-0.3V, $I_t$=500pA, $T$=4.2 K). The triangle-shaped dark features are from the Cr atoms embedded in the (Bi,Sb)$_2$Te$_3$ matrix. (**C**) $m$ dependence of the $dI/dV$ spectra on the surface of the ($m$,38) heterostructure (setpoint: $V_B$=+10mV, $I_t$=300pA, $T$=310 mK). (**D**) $m$ dependence of the superconducting gap $\Delta$ on the surface of the ($m$,38) heterostructure. The values of the superconducting gap $\Delta$ are determined by fitting the $dI/dV$ spectra. The error bars are the results of the fitting process and spatial distribution. The dashed line is a guide to the eyes. (**E**) Temperature dependence of the $dI/dV$ spectra on the surface of the (8, 38) heterostructure (setpoint: $V_B$=+10mV, $I_t$=300pA). The red dashed lines in (**C** and **E**) correspond to the zero $dI/dV$ values of the (1, 38) and (8, 38) heterostructures, respectively.

14